\documentclass{article}
\usepackage{spconf,amsmath,graphicx}
\usepackage[utf8]{inputenc} 
\usepackage[T1]{fontenc}    
\usepackage{hyperref}       
\usepackage{url}            
\usepackage{booktabs}       
\usepackage{amsfonts}       
\usepackage{nicefrac}       
\usepackage{microtype}      
\usepackage{xcolor}         
\usepackage{graphicx}
\usepackage{wrapfig}
\usepackage{multirow}

\title{Time Domain Adversarial Voice Conversion for ADD 2022}
\name{Cheng Wen, Tingwei Guo, Xingjun Tan, Rui Yan, Shuran Zhou, Chuandong Xie, Wei Zou, Xiangang Li}
\address{Beike, Beijing, China}
\email{\{wencheng008, zouwei026,lixiangang002\}@ke.com}

\begin{document}
\maketitle

\begin{abstract}
In this paper, we describe our speech generation system for the first Audio Deep Synthesis Detection Challenge (ADD 2022). Firstly, we build an any-to-many voice conversion (VC) system to convert source speech with arbitrary language content into target speaker’s fake speech. Then the converted speech generated from VC is post-processed in time-domain to improve the deception ability. The experimental results show that our system has adversarial ability against anti-spoofing detectors with a little compromise in audio quality and speaker similarity. This system ranks top in Track 3.1 in the ADD 2022, showing that our method could also gain good generalization ability against different detectors.
\end{abstract}
\section{Introduction}
\label{sec:intro}
Thanks to the superiority of deep learning and large-scale of high-quality open source speech corpus \cite{shi2021aishell} \cite{guo2021didispeech}, computational generated speech has reached humanlike naturalness and hi-fidelity audio quality. With a small batch of recording audio samples, state-of-art synthesis systems can generate non-distinguishable speech with high similarity of the target speaker. However, these technologies threaten the anti-spoofing and automatic speaker verification (ASV) systems greatly. In a recent study \cite{muller2021human}, synthetic speech is perceptually non-distinguishable from bona fide speech, and even well trained human detectors can get only 80\% in accuracy.

According to a survey \cite{khanjani2021deep}, audio deepfake methods can divide into three subcategories: replay attack, speech synthesis and voice conversion.
Replay attacks are defined as replaying the recording of a target speaker’s voice. Although the method is simple and efficient, it’s application is constrained by recording environment and language content. 
Speech synthesis (SS), also know as text to speech (TTS), is a technique that convert written language into human speech. Neural network-based SS systems can generate deepfake audios with a significant improvement in both intelligibility and naturalness, especially those with the end-to-end architectures, such as \cite{shen2018natural}, \cite{ping2017deep}, \cite{ren2019fastspeech}, \cite{kumar2019melgan}, \cite{kong2020hifi}. The main benefit of SS is that it can generate speech with arbitrary language content. Another benefit of SS is that it can generate any speaker’s voice with the development of adaptive TTS \cite{tan2021survey} technology, such as \cite{arik2018neural} and \cite{chen2021adaspeech}.
Voice conversion (VC) is a technique that converts a source speaker’s voice to a target speaker’s voice without changing linguistic information. The latest out-of-standing VC systems trend to utilize Variational Autoencoder (VAEs) and Generative Adversarial Network (GAN) frameworks to improve the target speaker similarity and audio quality. Such as Cycle-VAE \cite{tobing2019non}, Disentangled-VAE \cite{luong2021many}, fang’s CycleGAN-based nonparallel VC \cite{fang2018high}, STARGAN-VC \cite{kameoka2018stargan} and StarGANv2-VC \cite{li2021starganv2}.

Aim to accelerate and foster research on detecting deep synthesis and manipulated audios, Audio Deep Synthesis Detection Challenge \cite{Yi2022ADD} is held as Signal Processing Grand Challenge on ICASSP 2022. The challenge contains four tracks, among which the Track 3 is an audio fake game (FG) which includes two sub tasks: Track 3.1 Generation task (FG-G) is a generation task aims to generate fake audios that can fool the fake detection model. Track 3.2 Detection task (FG-D) is a detection task tries to detect all the generated fake audios, including results from FG-G. Our team has participated in the FG-G task and won the top rank.

This paper describes our contributions about audio deepfake anti-spoofing, especially VC-based fake speech generation method to “attack” the neural network-based detection systems. The backbone of our system is FastSpeech-VC \cite{zhao2021towards} followed with HiFi-GAN \cite{kong2020hifi}. Our FastSpeech- VC is designed to converts bottleneck feature (BNF) into mel- spectrograms. And then generate audio signal from the mel- spectrograms by the HiFi-GAN. In order to fool the detection systems, we further add a post-processing modification on the generated audio, which cause a slight decay in audio quality but a significant promotion in spoofing. Audio samples are available at our demo page\footnote{https://guo-t-w.github.io/KeAI-ADD-2022/}

The rest of this paper is organized as follows. Section 2 introduces our proposed method. Section 3 describes our implementation details and experimental results. Finally, the conclusion is given in section 4.

\section{SYSTEM DESCRIPTION}
\label{sec:format}
The goal of the Track 3.1 FG-G is to generate fake audio that can fool the detection system, which requires the distribution overlap between the generated audio and target speaker's natural audio to be as high as possible.

\begin{figure}[!h]
  \centering
  \includegraphics[trim = 0mm 0mm 0mm 0mm, width=0.52\textwidth]{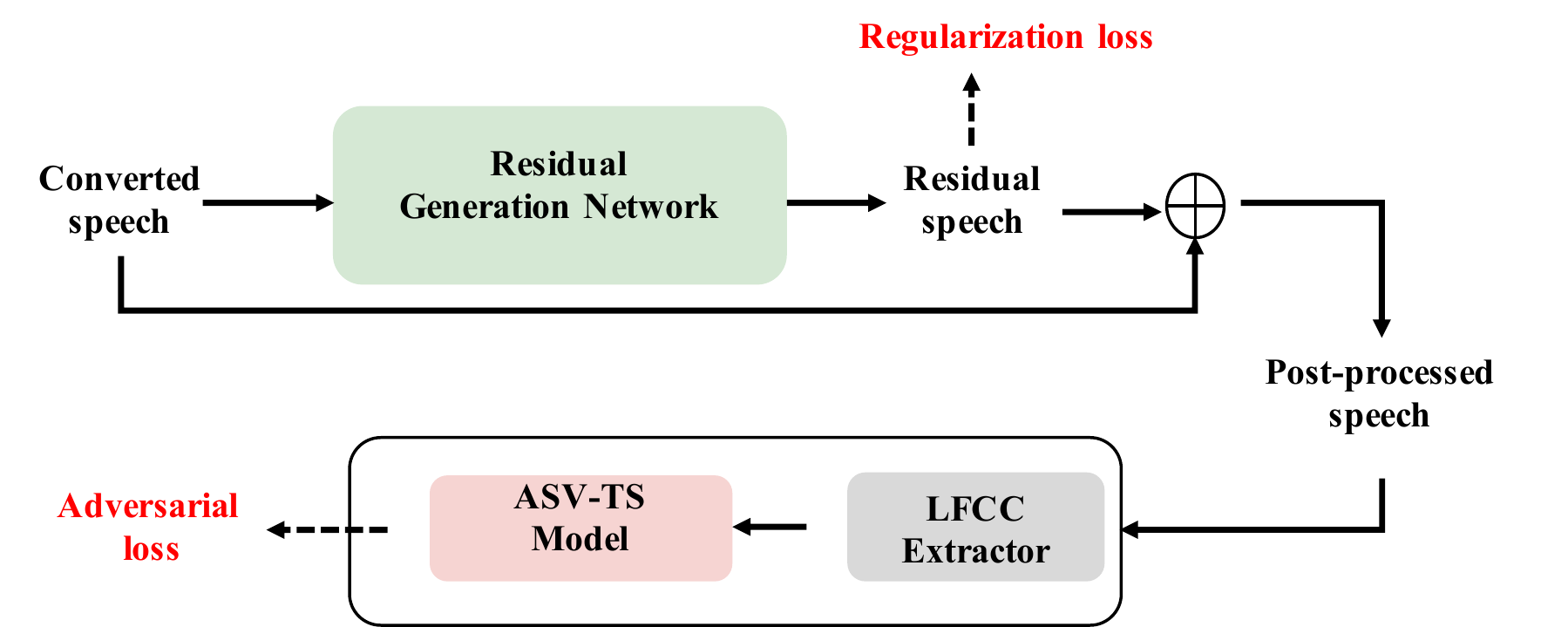}
  \caption{Time-Domain Adversarial Post-Processing framework}
  \label{fig:APN-framework}
\end{figure}

Inspired by \cite{ding2021adversarial}, we propose an adversarial approach to make the distribution of the VC generated converted speech, to be closer to that of the natural audio of target speaker.
As shown in Figure~\ref{fig:APN-framework}, the converted speech is sent to a residual generation network (RGN) to enable a white-box attack on a pretrained detection model.

Different from \cite{ding2021adversarial}, we build an automatic speaker verification system for targeted speakers (ASV-TS), which is used as the anti-spoofing system for the adversarial training of the residual generation network.
Given the set of all speech data $D=\{D_T, D_O\}$, where $D_T$ is the natural audio set of target speakers and $D_O$ denotes the set of all other speech data, the goal of ASV-TS is to distinguish $D_T$ from $D_O$ as accurately as possible.
However, the separating hyperplane between $D_T$ and $D_O$ determined by the ASV-TS may not perform well in complex scenarios, since only limited audios of target speakers in AIShell-3 could be used in Track 3.1 in the ADD 2022. 
In order to improve the performance of the ASV-TS system, various data augmentation techniques is adopted on the available speech data in $D_O$. In addition, the residual signal is generated based on the time-domain signal directly, instead of the spectral amplitude, to avoid loss of information.

Our entire system involves two stages, i.e., voice conversion and time-domain adversarial post-processing, which will be introduced in detail next.

\subsection{Voice Conversion}
\begin{figure*}[!h]
  \centering
  \includegraphics[trim = 0mm 0mm 0mm 0mm, width=0.8\textwidth]{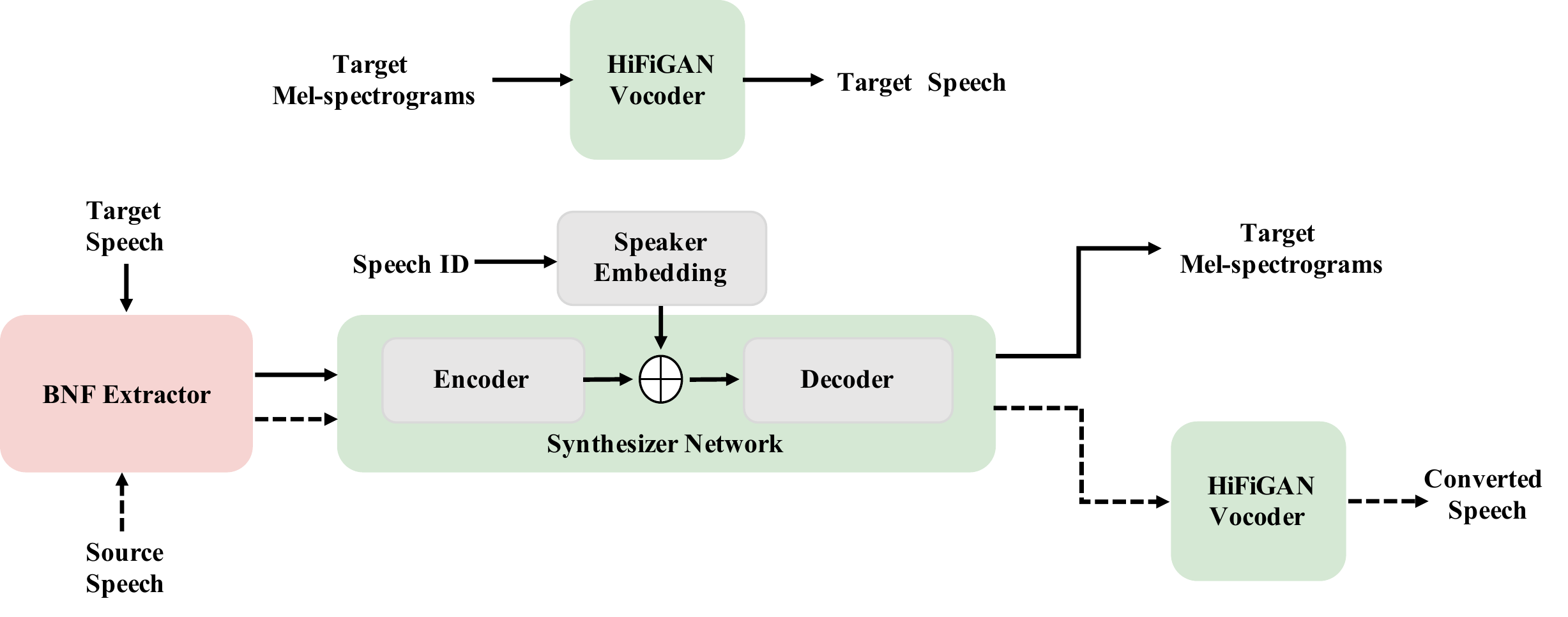}
  \caption{Voice Conversion framework.The solid and dashed lines represent the data flow for training and inference respectively}
  \label{fig:vc-framework}
\end{figure*}

As shown in Figure~\ref{fig:vc-framework}, a VC system is built to generate audio for specific timbres and content. The overall framework consists of three parts, BNF extractor, synthesizer network and vocoder.

Both phonetic posteriorgram (PPG) and BNF are widely used features in current VC tasks, which retain acoustic information while excluding speaker identities and are obtained from the ASR acoustic model. In our work, we use a DNN-based acoustic model from an ASR system trained with our own data, to perform as the BNF extractor, which consists of 7 TDNN layers and 3 unidirectional LSTM layers, followed by the 512-dim bottleneck layer.

Our synthesizer network is based on Fastspeech-VC framework \cite{zhao2021towards}, which originates from FastSpeech \cite{ren2019fastspeech} TTS model. The synthesizer network can predict mel-spectrograms from BNF. Meanwhile, speaker ids are used to control the speaker identity of synthesized utterances. Therefore, our system can achieve any to many voice conversions without reliance on the training data of source speakers. In the synthesizer network, the encoder and the decoder are composed of a stack of N = 6 duplicate Feed-Forward Transformer (FFT) blocks.

Finally, HiFi-GAN vocoder is used to reconstruct audio from  predicted mel-spectrograms.

\subsection{Time-Domain Adversarial Post-Processing}
\subsubsection{Target Speaker Verification Model}
Commonly used ResNet-34 \cite{he2016deep} is adopted to build the ASV-TS model in our work. The loss function is binary cross-entropy. We used linear frequency cepstrum coefficients (LFCCs) with a window size of 25ms and an overlap of 10ms as input of detection  model.
\subsubsection{Residual Generation Network}
The RGN in our system is a fully convolutional feed-forward network with waveform as input. It's architecture is similar to the generator of MelGAN \cite{kumar2019melgan}. Residual signal is generated by RGN and are then added to the input audio to obtain the post-processed speech. 
\subsubsection{Adversarial Training}
Given $s\in R^T$ denote the samples of input signal, the residual signal generated by RGN can be defined as $P(s)\in R^T$. When training the RGN, the adversarial training is conducted with objectives as:
\begin{equation}
L_A=1-D_t(F(s+P(s)))  
\end{equation}
where F is the LFCC features extractor and $D_t(.)$ represents the probability of the target audio predicted by the ASV-TS model. In addition, a regularization loss, which is composed of three parts, is also designed to maintain the subjective quality of post-processed speech.
\begin{equation}
L_R=L_r+ L_m+ L_s
\end{equation}
Among them, the $L_r$ is the difference between the maximum and the minimum sample values in the generated residual waveform. The $L_m$ is the mean square error (MSE) between the value of generated residual samples and a zero vector. By denoting the post-processing modification at each sample as
\begin{equation}
M_t=\frac{abs(P(s_t))}{abs(P(s_t))+abs(s_t)+0.0001}
\end{equation}
where $P(s_t)$ and $s_t$ are the generated residual and input values of t-th sample. The $L_s$ is defined as 
\begin{equation}
L_s=\frac{1}{T}\sum_{t=1}^TM_t
\end{equation}
The $L_s$ can further guarantee that the post-processing modifications are as slight as possible, especially in silence clips.

The final loss function of our post-processing model is
\begin{equation}
L=\lambda_A L_A + \lambda_R L_R
\end{equation}
where $\lambda_A$ and $\lambda_R$ are the scale factors of adversarial and regularization loss respectively.

\section{EXPERIMENTS}
\label{sec:exp}

\subsection{Datasets and Model Details}

\begin{itemize}
\item[$\bullet$] \textbf{Voice Conversion}: For Track3.1 Generation task, participants are required to generate deepfake audio of 10 target speakers using the AIShell-3 dataset \cite{shi2021aishell}. We first trained multi-speaker synthesizer network and universal vocoder based on the AIShell-3, and then fine-tuned them for all target speakers. After that, 5000 deepfake audios of 10 target speakers, denoted as VC-T5000, are generated. The deepfake waveform of each speaker is produced with 500 utterances recorded by mobile phone in quiet  environment as source speech.
\item[$\bullet$] \textbf{ASV-TS Model}: ASV-TS model is trained to distinguish the natural speech of target speakers from other audios. We take all utterances of 10 target speakers as positive samples. Meanwhile, the same amount of audios of other speakers randomly selected from AIShell-3 and deepfake audios generated from VC and TTS (built on AIShell-3) are used as negative audios. In addition, each of negative audio is augmented by a method randomly chosen from Table~\ref{table:list of data aug} to further extend the set of negative audios.
\item[$\bullet$] \textbf{RGN}: For the training of RGN, all audios in VC-T5000 are used, and the ASV-TS model is employed to perform adversarial training. 
In the post-processing network, four upsampling layers with 8x, 8x, 2x, 2x factors respectively are used to achieve 256x upsampling. All input audios are reshaped with shape [T, 256], and the shape of output residual signal is [T*256, 1]. The values of learning rate  are [0.0001, 0.00005, 0.000025, 0.0000125, 0.00000625], where decay boundaries are [5000, 10000, 30000, 50000] steps. In order to guarantee the post-processing modifications are as slight as possible, we set $\lambda_R=20$ and $\lambda_A=1$.
\end{itemize}

\begin{table}[!h]
  \caption{List of Data Augmentation Methods}
  \centering
  \resizebox{0.5\textwidth}{!}{
  \begin{tabular}{llc}
    \toprule
   \begin{tabular}{ccccc}
   \multicolumn{1}{c}{Approach}
    & \multicolumn{1}{c}{Methods}
    & \multicolumn{1}{c}{Description} \\
    \midrule
    \multirow{5}{*}{Distortion} & noise & MUSAN and \\
    & music & self-collected\\
    & babble & \\ 
    \cline{2-3}
    & reverb & room impulse response \\
    \cline{2-3}
    & volume & -10dB to 20dB \\
    \midrule
    \multirow{5}{*}{Compression} & MP3 & & \\
    & OGG & Random\\
    & AAC & compression ratio\\
    & OPUS \\
    \cline{2-3}
    & sample rate & 16kHz -> 8kHz\\
\bottomrule
\label{table:list of data aug}
 \end{tabular}
\end{tabular}
}
\end{table}

\subsection{Evaluation}
The deception success rate (DSR) is chosen as the metric for generation task. DSR is defined as followed:
\begin{equation}
DSR=\frac{W}{A * N}
\end{equation} 
where W is the count of wrong detection samples by all the detection models on the condition of reaching each own EER performance, A is the count of all the evaluation samples, and N is the number of detection models. 

\begin{figure}[!h]
  \centering
  \includegraphics[trim = 0mm 0mm 0mm 0mm, width=0.5\textwidth]{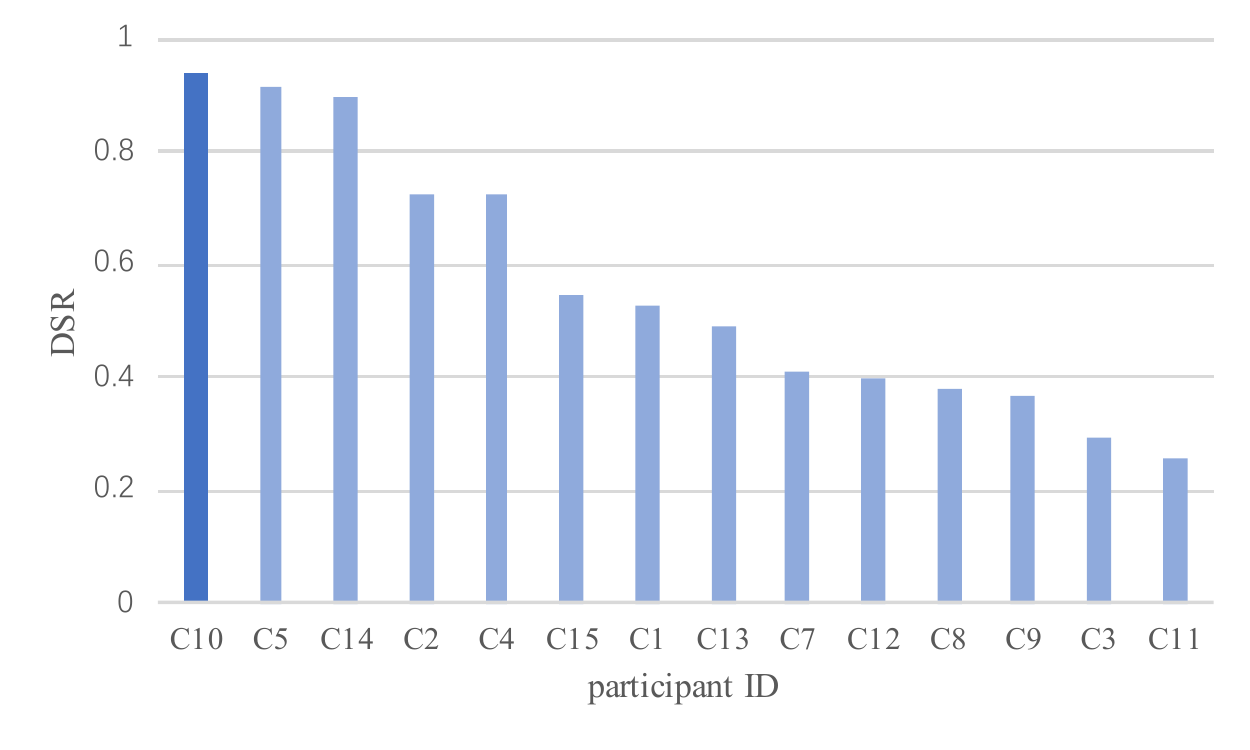}
  \caption{The final DSR score of our submited audios}
  \label{fig:dsr}
\end{figure}

The results are shown in Figure~\ref{fig:dsr}. Our team id is C10. It can be seen that our final result ranks first, which also shows the effectiveness of our proposed model.

\subsection{Analysis}
We analysed the character error rate (CER) and the cosine similarity (COS\_SIM) of speaker embedding vectors of the speech before and after post-processing. As shown in Table~\ref{table:cer}, the CER and cosine similarity have no significant change, showing that post-processing has no effect on audio quality.

In order to compare the spoofing performance of before and after post-processing, a new spoofing detection model is trained based on the data of ADD 2022 Detection Task. The model configurations are the same as ASV-TS model.
We randomly select 1.4K utterances from the audio before and after post-processing, respectively, and combine them with 0.6K utterances from  AIShell-3 as two test sets. From Table~\ref{table:cer-1} we can find that, the EER changed from 24.7\% to 74.0\%, which means the deception ability of post-processed speech is significantly improved.

\begin{table}[!h] 
  \caption{The CER and cosine similarity of speaker embedding vectors the speech before and after postprocessing}
  \centering
  \resizebox{0.5\textwidth}{!}{
  \begin{tabular}{llc}
    \toprule
   \begin{tabular}{cccccc}
   \multicolumn{1}{c}{Speaker ID}
    & \multicolumn{1}{c}{Source}
    & \multicolumn{2}{c}{Before}  
    &\multicolumn{2}{c}{After}\\ 
    \cmidrule(r){2-2} \cmidrule(r){3-4} \cmidrule(r){5-6}
    &CER(\%) & CER(\%) & COS\_SIM & CER(\%) & COS\_SIM \\
    \midrule
    SSB0139      & \_       &  14.39   & 0.93 & 14.76 & 0.91\\
    SSB0535      & \_       &  13.24   & 0.91 & 13.26 & 0.90\\
    SSB0601      & \_       &  13.70   & 0.90 & 14.46 & 0.88\\ 
    SSB0603      & \_       &  13.63   & 0.90 & 13.86 & 0.87\\
    SSB0607      & \_       &  13.27   & 0.89 & 14.26 & 0.85\\ 
    SSB0609      & \_       &  13.96   & 0.89 & 14.75 & 0.86\\
    SSB0629      & \_       &  13.83   & 0.94 & 14.04 & 0.91\\
    SSB0666      & \_       &  13.76   & 0.91 & 14.68 & 0.89\\
    SSB0668      & \_       &  13.13   & 0.93 & 13.49 & 0.91\\
    SSB0671      & \_       &  13.54   & 0.87 & 14.02 & 0.84\\
    \midrule
    Average      & 10.41    &  13.65   & 0.91 & 14.16 & 0.88\\
\bottomrule
\label{table:cer}
 \end{tabular}
\end{tabular}
}
\end{table}

\begin{table}[!h]
\vspace{-0.1em}
  \caption{The EER of the speech before and after postprocessing}
  \centering
  \resizebox{0.2\textwidth}{!}{

  \begin{tabular}{llc}
    \toprule
        &   Before  &  After \\
    \midrule
    EER         & 24.7\%   &   74.0\% \\
\bottomrule
\label{table:cer-1}
\end{tabular}
  }
\vspace{-0.1em}
\end{table}

\begin{figure}[!h]
  \centering
  \includegraphics[trim = 0mm 0mm 0mm 0mm, width=0.45\textwidth, height=0.3\textwidth]{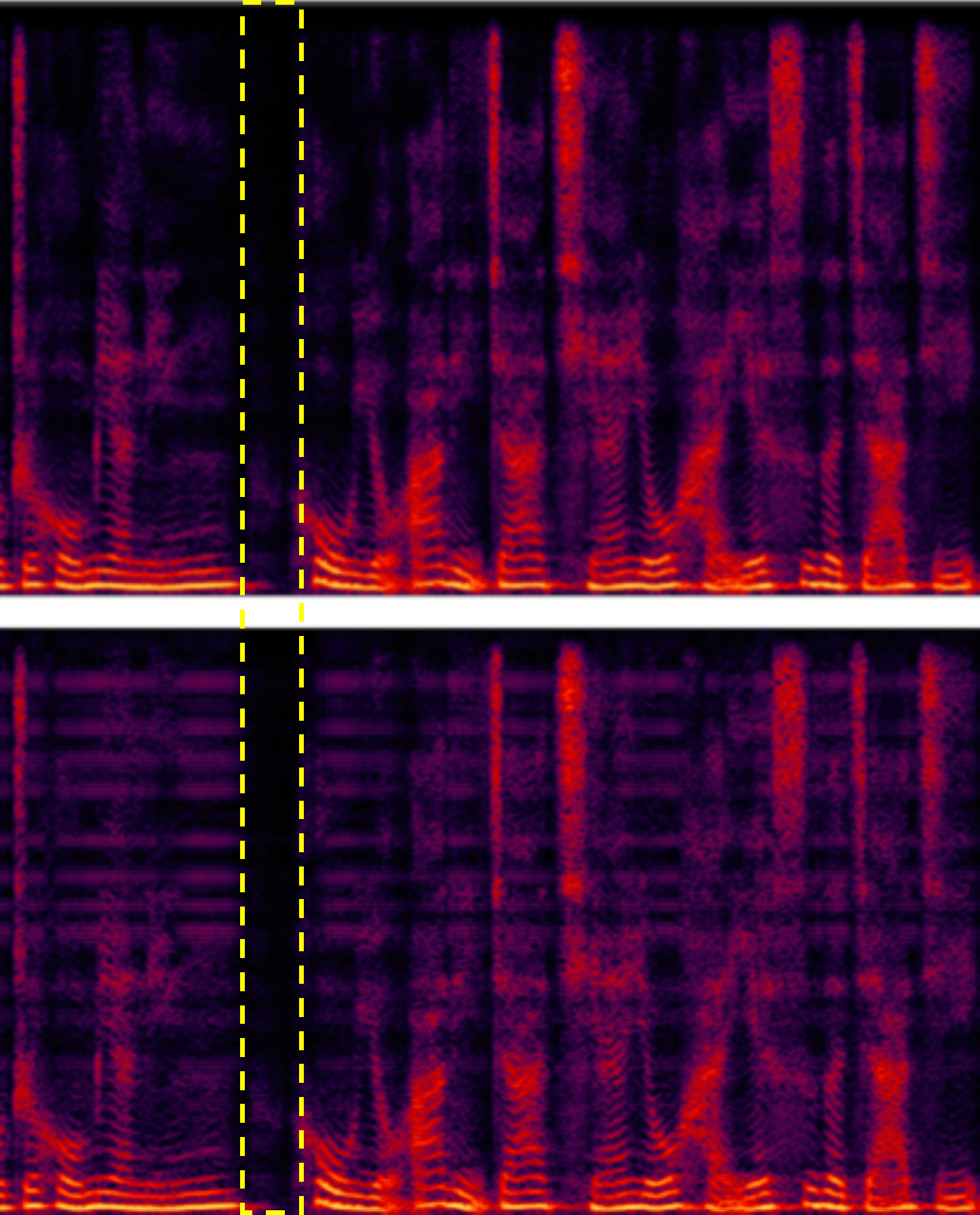}
  \caption{The spectrograms of the before (top) and after audio (bottom) post-processing}
  \label{fig:spec}
\end{figure}

Furthermore, we analyzed the spectrum of speech. Figure \ref{fig:spec} illustrates the spectrograms before and after post-processed by RGN model. It can be seen that the difference between the two spectrograms is very slight, especially in low frequency region. In addition, there is nearly no residual signal has been added to the silence clips in the input audio, which means RGN only perturbs where there is speech information.

\section{CONCLUSION}
\label{sec:typestyle}
This paper introduces our system for FG-G. It involves two stages, including voice conversion and time-domain adversarial post-processing. The voice conversion model is built using Fastspeech-VC framework and the post-processing is performed directly on the time-domain of converting audios by the RGN, which is implemented by adversarial training against the ASV-TS. The experimental results show that our system can generate audios with both high quality and adversarial ability against spoofing detectors. Our system has also achieved the highest performance, a DSR of 0.938, among all participants of the Track 3.1 of Audio Deep Synthesis Detection Challenge 2022.

\clearpage


\bibliographystyle{IEEEbib}
\bibliography{strings,refs}

\begin{thebibliography}{10}

\bibitem{shi2021aishell}
Yao Shi, Hui Bu, Xin Xu, Shaoji Zhang, and Ming Li,
\newblock ``Aishell-3: A multi-speaker mandarin tts corpus,''
\newblock {\em Proc. Interspeech 2021}, pp. 2756--2760, 2021.

\bibitem{guo2021didispeech}
Tingwei Guo, Cheng Wen, Dongwei Jiang, Ne~Luo, Ruixiong Zhang, Shuaijiang Zhao,
  Wubo Li, Cheng Gong, Wei Zou, Kun Han, et~al.,
\newblock ``Didispeech: A large scale mandarin speech corpus,''
\newblock in {\em ICASSP 2021-2021 IEEE International Conference on Acoustics,
  Speech and Signal Processing (ICASSP)}. IEEE, 2021, pp. 6968--6972.

\bibitem{muller2021human}
Nicolas~M M{\"u}ller, Karla Markert, and Konstantin B{\"o}ttinger,
\newblock ``Human perception of audio deepfakes,''
\newblock {\em arXiv preprint arXiv:2107.09667}, 2021.

\bibitem{khanjani2021deep}
Zahra Khanjani, Gabrielle Watson, and Vandana~P Janeja,
\newblock ``How deep are the fakes? focusing on audio deepfake: A survey,''
\newblock {\em arXiv preprint arXiv:2111.14203}, 2021.

\bibitem{shen2018natural}
Jonathan Shen, Ruoming Pang, Ron~J Weiss, Mike Schuster, Navdeep Jaitly,
  Zongheng Yang, Zhifeng Chen, Yu~Zhang, Yuxuan Wang, Rj~Skerrv-Ryan, et~al.,
\newblock ``Natural tts synthesis by conditioning wavenet on mel spectrogram
  predictions,''
\newblock in {\em 2018 IEEE International Conference on Acoustics, Speech and
  Signal Processing (ICASSP)}. IEEE, 2018, pp. 4779--4783.

\bibitem{ping2017deep}
Wei Ping, Kainan Peng, Andrew Gibiansky, Sercan~O Arik, Ajay Kannan, Sharan
  Narang, Jonathan Raiman, and John Miller,
\newblock ``Deep voice 3: Scaling text-to-speech with convolutional sequence
  learning,''
\newblock {\em arXiv preprint arXiv:1710.07654}, 2017.

\bibitem{ren2019fastspeech}
Yi~Ren, Yangjun Ruan, Xu~Tan, Tao Qin, Sheng Zhao, Zhou Zhao, and Tie-Yan Liu,
\newblock ``Fastspeech: Fast, robust and controllable text to speech,''
\newblock {\em arXiv preprint arXiv:1905.09263}, 2019.

\bibitem{kumar2019melgan}
Kundan Kumar, Rithesh Kumar, Thibault de~Boissiere, Lucas Gestin, Wei~Zhen
  Teoh, Jose Sotelo, Alexandre de~Br{\'e}bisson, Yoshua Bengio, and Aaron
  Courville,
\newblock ``Melgan: Generative adversarial networks for conditional waveform
  synthesis,''
\newblock {\em arXiv preprint arXiv:1910.06711}, 2019.

\bibitem{kong2020hifi}
Jungil Kong, Jaehyeon Kim, and Jaekyoung Bae,
\newblock ``Hifi-gan: Generative adversarial networks for efficient and high
  fidelity speech synthesis,''
\newblock {\em arXiv preprint arXiv:2010.05646}, 2020.

\bibitem{tan2021survey}
Xu~Tan, Tao Qin, Frank Soong, and Tie-Yan Liu,
\newblock ``A survey on neural speech synthesis,''
\newblock {\em arXiv preprint arXiv:2106.15561}, 2021.

\bibitem{arik2018neural}
Sercan~O Arik, Jitong Chen, Kainan Peng, Wei Ping, and Yanqi Zhou,
\newblock ``Neural voice cloning with a few samples,''
\newblock {\em arXiv preprint arXiv:1802.06006}, 2018.

\bibitem{chen2021adaspeech}
Mingjian Chen, Xu~Tan, Bohan Li, Yanqing Liu, Tao Qin, Sheng Zhao, and Tie-Yan
  Liu,
\newblock ``Adaspeech: Adaptive text to speech for custom voice,''
\newblock {\em arXiv preprint arXiv:2103.00993}, 2021.

\bibitem{tobing2019non}
Patrick~Lumban Tobing, Yi-Chiao Wu, Tomoki Hayashi, Kazuhiro Kobayashi, and
  Tomoki Toda,
\newblock ``Non-parallel voice conversion with cyclic variational
  autoencoder,''
\newblock {\em Proc. Interspeech 2019}, 2019.

\bibitem{luong2021many}
Manh Luong and Viet~Anh Tran,
\newblock ``Many-to-many voice conversion based feature disentanglement using
  variational autoencoder,''
\newblock {\em arXiv preprint arXiv:2107.06642}, 2021.

\bibitem{fang2018high}
Fuming Fang, Junichi Yamagishi, Isao Echizen, and Jaime Lorenzo-Trueba,
\newblock ``High-quality nonparallel voice conversion based on cycle-consistent
  adversarial network,''
\newblock in {\em 2018 IEEE International Conference on Acoustics, Speech and
  Signal Processing (ICASSP)}. IEEE, 2018, pp. 5279--5283.

\bibitem{kameoka2018stargan}
Hirokazu Kameoka, Takuhiro Kaneko, Kou Tanaka, and Nobukatsu Hojo,
\newblock ``Stargan-vc: Non-parallel many-to-many voice conversion using star
  generative adversarial networks,''
\newblock in {\em 2018 IEEE Spoken Language Technology Workshop (SLT)}. IEEE,
  2018, pp. 266--273.

\bibitem{li2021starganv2}
Yinghao~Aaron Li, Ali Zare, and Nima Mesgarani,
\newblock ``Starganv2-vc: A diverse, unsupervised, non-parallel framework for
  natural-sounding voice conversion,''
\newblock {\em arXiv preprint arXiv:2107.10394}, 2021.

\bibitem{Yi2022ADD}
Jiangyan Yi, Ruibo Fu, Jianhua Tao, Shuai Nie, Haoxin Ma, Chenglong Wang, Tao
  Wang, Zhengkun Tian, Ye~Bai, Shan Liang, Shiming Wang, Shuai Zhang, Xinrui
  Yan, Le~Xu, and Haizhou Li,
\newblock ``Add 2022: the first audio deep synthesis detection challenge,''
\newblock in {\em 2022 IEEE International Conference on Acoustics, Speech and
  Signal Processing (ICASSP)}. IEEE, 2022.

\bibitem{zhao2021towards}
Shengkui Zhao, Hao Wang, Trung~Hieu Nguyen, and Bin Ma,
\newblock ``Towards natural and controllable cross-lingual voice conversion
  based on neural tts model and phonetic posteriorgram,''
\newblock in {\em ICASSP 2021-2021 IEEE International Conference on Acoustics,
  Speech and Signal Processing (ICASSP)}. IEEE, 2021, pp. 5969--5973.

\bibitem{ding2021adversarial}
Yi-Yang Ding, Li-Juan Liu, Yu~Hu, and Zhen-Hua Ling,
\newblock ``Adversarial voice conversion against neural spoofing detectors,''
\newblock in {\em Proc. INTERSPEECH 2021}, 2021, pp. 816--820.

\bibitem{he2016deep}
Kaiming He, Xiangyu Zhang, Shaoqing Ren, and Jian Sun,
\newblock ``Deep residual learning for image recognition,''
\newblock in {\em Proceedings of the IEEE conference on computer vision and
  pattern recognition}, 2016, pp. 770--778.

\end{thebibliography}

\end{document}